# Decoding angular light paths for solving the inverse scattering problem

Kalpak Gupta[1], Dinh Hoang Tran[1], Sungsam Kang[1], Yongwoo Kwon[1], Seokchan Yoon[2], Jin Hee Hong[1], Ye-Ryoung Lee[3,*], and Wonshik Choi[1,*]

[1]*Department of Physics, Korea University, Seoul 02841, Korea*

[2]*School of Biomedical Convergence Engineering, Pusan National University, Yangsan 50612, Korea*

[3]*Department of Physics, Konkuk University, Seoul, 05029, Korea*

[*]*Corresponding author. E-mail: yeryoung@konkuk.ac.kr, wonshik@korea.ac.kr*

**Abstract**

Multiple scattering in complex media scrambles the deterministic mapping between input and output fields, limiting wave control and imaging. Conventional inverse scattering strategies rely on discrete spatial layers, but this assumption breaks down in volumetric media where scattering is continuously distributed, particularly near the object plane. Here we introduce a framework that reformulates light transport in terms of scattering angles rather than spatial layers. We show that decomposing scattered waves into angular deflection components—each associated with a spatially invariant point spread function—provides a compact and depth-independent description of volumetric scattering. This representation is particularly effective in forward-scattering biological tissues, where most scattered energy is confined to a narrow angular range. Leveraging this angular sparsity, we develop a progressive inverse algorithm that retrieves dominant angular components from reflection measurements, converting multiply scattered light into usable signal by more than an order of magnitude of the ballistic signal. We demonstrate *in vivo* recovery of subcellular osteocyte networks through intact mouse skulls, a regime inaccessible to existing methods. These results establish the scattering-angle basis as a general physical framework for decoding information scrambled by disorder, extending the reach of deep optical imaging.

**Introduction**

Wave propagation in disordered media is a central problem in optics and acoustics, where multiple scattering scrambles the deterministic mapping between incident and transmitted fields[1–3]. These scattering events randomize both the phase and direction of waves, hindering our ability to control and invert wave transport. This limitation constrains not only fundamental studies of complex systems but also a wide range of applications, including optical microscopy, communication through turbid media, and ultrasonic imaging in heterogeneous environments. In the context of high-resolution optical microscopy[1], the primary challenge is to identify and correct the complex distortions a focused beam experiences as it traverses scattering tissue. This distortion, characterized by the point spread function (PSF), degrades image resolution and contrast. The problem is particularly severe in reflection geometries, essential for *in vivo* imaging, where the wavefront is distorted along both the forward and return paths[4].

Numerous approaches have been developed to address the inverse scattering problem, including wavefront shaping[2,3,5,6], speckle correlations[7,8], and computational reconstruction[9,10]. The majority of these methods rely on the assumption that scattering originates from a dominant layer well-separated from the object. Under this condition, the PSF, while distorted, remains approximately invariant with respect to focus scanning over a localized region known as the isoplanatic patch[11–14]. This spatial invariance forms the basis for conventional aberration correction. However, this assumption breaks down in volumetric scattering media, where scatterers are distributed continuously throughout the sample, including the depths near the target object. In this regime, light reaching the focal plane spans a broad range of angles, causing the PSF to vary rapidly across space, thereby violating the shift-invariance required for correction. While patch-wise reconstruction[9,15–17] attempts to mitigate this by segmenting the field of view, it fails when the PSF variation becomes too steep. Conjugate adaptive optics[18–20] and multiconjugate approaches[21–23] attempt to address this by placing the correction layers conjugate to the scattering medium, but these are effective only when dominant scatterers are located at a distance from the object. Accounting for scattering near the object plane would require an impractically large number of closely spaced correction layers, rendering the inverse problem highly ill-posed.

To overcome these limitations, we introduce a framework that reformulates wave transport in terms of angular deflections rather than spatial layers. Instead of discretizing the medium into localized layers, we represent the scattering process as a coherent superposition of angular modes, where each mode corresponds to a specific scattering angle. In this *scattering-angle basis*, we identify a crucial physical property: every angular component contributes a shift-invariant PSF modulated by a simple deterministic phase factor. The complex, spatially varying PSF of the volumetric medium then emerges from the interference of these angular components. This formulation not only generalizes prior statistical models relating angular spread to isoplanatic patch size[24] but also offers a compact physical description that decouples the scattering model from the specific depths of scatterers. Furthermore, in forward-scattering media such as biological tissue, energy is confined to relatively small deflection angles, allowing the medium's response to be described with a finite number of angular components.

Building on this principle, we present a progressive inverse algorithm that recovers the dominant scattering-angle components directly from reflection-mode measurements. By iteratively expanding the range of reconstructed scattering angles, our method converts multiply scattered signals traditionally discarded as noise into usable information, enabling the correction of steep, position-dependent PSFs. We validate this framework by resolving subcellular structures, specifically osteocytes and their dendritic processes, within intact bone tissue, a regime effectively inaccessible to conventional methods due to strong scattering near the object plane. By establishing a deterministic link between angular scattering modes and image formation, our work provides a general physical framework for understanding and inverting wave propagation in strongly scattering environments.

**The scattering-angle basis for describing spatially varying point spread function**

The angular spread of waves as they propagate through a scattering medium gives rise to spatial variations in the point spread function (PSF)[24]. Here we introduce a framework that describes how each scattering-angle component contributes an individual PSF, and how their superposition determines the overall, spatially varying PSF. In this formulation, each transmitted wave is characterized by the deflection angle it experiences upon scattering. For instance, if a wave propagates without any change in direction, the deflection is zero, i.e., $\Delta\mathbf{k} = \mathbf{0}$, where $\Delta\mathbf{k} \equiv \mathbf{k} - \mathbf{k}_\mathrm{i}$ is the difference between the transverse wavevector of the transmitted wave ($\mathbf{k}$) and that of the incident wave ($\mathbf{k}_\mathrm{i}$) (Fig. 1a). A wave that is scattered by a specific angle $\Delta\mathbf{k} = \Delta\mathbf{k}_\mathrm{s}$ is illustrated in Fig. 1e. The connection between input and output angles is described by the transfer function $\tilde{P}(\mathbf{k}; \mathbf{k}_\mathrm{i})$. Representing this relation in the angle-to-angle transmission matrix $\tilde{\boldsymbol{P}}$ naturally organizes angular deflections: undeviated (ballistic) waves lie along the main diagonal of $\tilde{\boldsymbol{P}}$ (Fig. 1b), while scattered waves with deflection $\Delta\mathbf{k}_\mathrm{s}$ correspond to a specific off-diagonal (Fig. 1f).

When we send a focused illumination, which is formed by the incident waves with various propagation angles, each deflection angle produces a distinct PSF. The PSF $P(\mathbf{r}; \mathbf{r}_\mathrm{i})$ is related to the transfer function $\tilde{P}(\mathbf{k}; \mathbf{k}_\mathrm{i})$ through the Fourier transform relationship (see Supplementary Note 1):

$$P(\mathbf{r}; \mathbf{r}_\mathrm{i}) = \int\int e^{i\mathbf{k}\cdot\mathbf{r}} \tilde{P}(\mathbf{k}; \mathbf{k}_\mathrm{i}) e^{-i\mathbf{k}_\mathrm{i}\cdot\mathbf{r}_\mathrm{i}}\, d\mathbf{k}\, d\mathbf{k}_\mathrm{i}. \quad (1)$$

Here $\mathbf{r}$ is the observation point and $\mathbf{r}_\mathrm{i}$ is the illumination position. The PSF formed by the ballistic waves ($\Delta\mathbf{k} = \mathbf{0}$) is spatially invariant, meaning it remains unchanged as the focus is scanned (Figs. 1c,d). We denote this PSF as $P^{(\Delta\mathbf{k}=\mathbf{0})}(\mathbf{r}; \mathbf{r}_\mathrm{i}) = W^{(\Delta\mathbf{k}=\mathbf{0})}(\mathbf{r} - \mathbf{r}_\mathrm{i})$.

In contrast, waves scattered at a non-zero deflection angle $\Delta\mathbf{k}_\mathrm{s}$ produce a PSF that varies with focus scanning (Figs. 1g,h). However, our theoretical derivation (see Supplementary Notes 1 and 2) reveals that this PSF is spatially invariant in amplitude, with variation only in the phase. Specifically, the PSF is given by:

$$P^{(\Delta\mathbf{k}_\mathrm{s})}(\mathbf{r}; \mathbf{r}_\mathrm{i}) = W^{(\Delta\mathbf{k}_\mathrm{s})}(\mathbf{r} - \mathbf{r}_\mathrm{i})\, e^{i\Delta\mathbf{k}_\mathrm{s}\cdot\mathbf{r}}, \quad (2)$$

where $W^{(\Delta\mathbf{k}_\mathrm{s})}(\mathbf{r} - \mathbf{r}_\mathrm{i})$ is a shift-invariant envelope, and the phase term $e^{i\Delta\mathbf{k}_\mathrm{s}\cdot\mathbf{r}}$ introduces spatially varying phase depending on the deflection angle (Fig. 1h). Importantly, the spatial phase modulation frequency increases with larger scattering angles.

If the scattering medium produces both ballistic and deflected waves by a specific angle $\Delta\mathbf{k}_\mathrm{s}$ (Fig. 1i), the matrix $\tilde{\boldsymbol{P}}$ contains both a diagonal and an off-diagonal term (Fig. 1j). The resulting PSF is a coherent superposition of the two components, i.e. $P(\mathbf{r}; \mathbf{r}_\mathrm{i}) = W^{(\Delta\mathbf{k}=0)} + W^{(\Delta\mathbf{k}_\mathrm{s})} e^{i\Delta\mathbf{k}_\mathrm{s}\cdot\mathbf{r}}$. This superposition leads to spatial variation in both amplitude and phase due to interference, as illustrated in Fig. 1k.

In a general scattering medium, light is scattered over a broad range of angles. The total PSF is then described as an integral over all deflection components:

$$P(\mathbf{r}; \mathbf{r}_\mathrm{i}) = \int W^{(\Delta\mathbf{k})}(\mathbf{r} - \mathbf{r}_\mathrm{i})\, e^{i\Delta\mathbf{k}\cdot\mathbf{r}} d\Delta\mathbf{k}. \quad (3)$$

Equation (3) expresses the PSF in the scattering-angle basis. As more deflection components are included, the accumulated phase terms introduce increasingly rapid spatial variations, resulting in a highly position-dependent PSF.

**Isoplanatic patch in the scattering-angle basis**

Our scattering-angle framework provides a physical origin of a finite isoplanatic patch size. The patch size can be measured by the translational correlation of the spatially varying PSF, which is defined as the normalized correlation between the PSF, $P(\mathbf{r};\mathbf{r}_\mathrm{i})$, at position $\mathbf{r}_\mathrm{i}$ and the shifted PSF, $P(\mathbf{r}+\Delta\mathbf{r}_\mathrm{s};\mathbf{r}_\mathrm{i}+\Delta\mathbf{r}_\mathrm{s})$, scanned by $\Delta\mathbf{r}_\mathrm{s}$ relative to $\mathbf{r}_\mathrm{i}$. The correlation function $C_\mathrm{i}(\Delta\mathbf{r}_\mathrm{s})$ can be written as (see Supplementary Note 3 for derivation):

$$C_\mathrm{i}(\Delta\mathbf{r}_\mathrm{s}) \propto \int\int \left|W_\mathrm{i}^{(\Delta\mathbf{k})}(\mathbf{r}-\mathbf{r}_\mathrm{i})\right|^2 e^{-i\Delta\mathbf{k}\cdot\Delta\mathbf{r}_\mathrm{s}}\, d(\mathbf{r}-\mathbf{r}_\mathrm{i})\, d\Delta\mathbf{k}\,. \quad (4)$$

We first consider the case of a single deflection angle $\Delta\mathbf{k}_\mathrm{s}$ (Figs. 2a,b). Because the PSF in Eq. (2) contains the scattering angle-dependent phase term $e^{i\Delta\mathbf{k}_\mathrm{s}\cdot\mathbf{r}}$, the resulting correlation oscillates with spatial frequency $\Delta\mathbf{k}_\mathrm{s}$. Hence, larger deflection angles produce faster spatial oscillations in the correlation function (Fig. 2b compared to Fig. 2a).

As the number of scattering angles increases, PSFs corresponding to different $\Delta\mathbf{k}$ values combine to form the total, spatially varying PSF. Consequently, their translational correlations also superpose. While all correlations are in phase at $\Delta\mathbf{r}_\mathrm{s}=0$, they acquire different phase shifts with increasing $\Delta\mathbf{r}_\mathrm{s}$, leading to destructive interference among the various $\Delta\mathbf{k}$ components. The resulting decay in the total correlation with $\Delta\mathbf{r}_\mathrm{s}$ quantitatively describes the finite isoplanatic patch size, consistent with prior statistical interpretations that relate it to the angular spread of scattering[24].

As an example, consider a scattered field whose angular amplitude spectrum follows a Gaussian distribution with standard deviation $\sigma_k$. In this case, the correlation function also follows a Gaussian envelope (Figs. 2c,d), with standard deviation $\sqrt{2}/\sigma_k$(see Supplementary Note 3). Thus, a broader angular spread produces a more rapidly decorrelating PSF, corresponding to a smaller isoplanatic patch.

**Solving the inverse problem in the scattering-angle basis**

Having established the theoretical foundation of spatially varying PSFs in the scattering-angle basis, we now describe how this framework can be applied for *in vivo* imaging. Reflection-mode geometry is essential for *in vivo* applications, since transmission-mode detection would require placing detectors within the tissue. However, this configuration is particularly challenging because the scattering medium induces angular deflections on both the incident and reflected waves (Fig. 3a). Four wavevectors are involved in the scattering processes: $\mathbf{k}_\mathrm{i}$ and $\mathbf{k}_\mathrm{o}$ for the incident and reflected fields, and $\mathbf{k}$ and $\mathbf{k}'$ for the waves arriving at and departing from the object.

In reflection-mode imaging, a series of complex-field images $\tilde{E}(\mathbf{k}_\mathrm{o};\mathbf{k}_\mathrm{i})$ is recorded at the detector plane for multiple incident wavevectors $\mathbf{k}_\mathrm{i}$ (Fig. 3b). These measurements form the reflection matrix $\widetilde{\boldsymbol{R}}$, with $\mathbf{k}_\mathrm{o}$ and $\mathbf{k}_\mathrm{i}$ as row and column indices, respectively[9] (Fig. 3c). The reflection matrix can be decomposed into three components that describe sequential scattering processes: (i) the input transmission matrix $\widetilde{\boldsymbol{P}}_\mathrm{i}$, which represents the distortion of the incident light on its way to the object; (ii) the object matrix $\widetilde{\boldsymbol{O}}$, which describes reflection from the target; and (iii) the output transmission matrix $\widetilde{\boldsymbol{P}}_\mathrm{o}$, which accounts for scattering on the return path to the detector (see Supplementary Note 4 and Supplementary Fig. 3). The measured reflection matrix is therefore expressed as $\widetilde{\boldsymbol{R}}=\widetilde{\boldsymbol{P}}_\mathrm{o}\widetilde{\boldsymbol{O}}\,\widetilde{\boldsymbol{P}}_\mathrm{i}$.

The inverse problem entails estimating $\widetilde{\boldsymbol{P}}_\mathrm{i}$, $\widetilde{\boldsymbol{O}}$, and $\widetilde{\boldsymbol{P}}_\mathrm{o}$ from the measured $\widetilde{\boldsymbol{R}}$. We solve this efficiently by leveraging the scattering-angle basis, in which each transmission matrix is decomposed into angular components, with each component corresponding to a distinct off-diagonal. Specifically, we estimate the approximate matrices $\widehat{\boldsymbol{P}}_\mathrm{i}$, $\widehat{\boldsymbol{O}}$, and $\widehat{\boldsymbol{P}}_\mathrm{o}$ such that the predicted reflection matrix $\widehat{\boldsymbol{R}}=\widehat{\boldsymbol{P}}_\mathrm{o}\widehat{\boldsymbol{O}}\,\widehat{\boldsymbol{P}}_\mathrm{i}$ best matches the measured $\widetilde{\boldsymbol{R}}$. The loss function to be minimized is defined as

$$\mathcal{L}=-\mathrm{PC}\,[\,\widetilde{\boldsymbol{R}},\widehat{\boldsymbol{P}}_\mathrm{o}\,\widehat{\boldsymbol{O}}\,\widehat{\boldsymbol{P}}_\mathrm{i}\,], \quad (5)$$

where PC denotes the Pearson correlation coefficient, quantifying the linear similarity between the two matrices (see Methods).

Directly optimizing all matrix elements would make the problem severely under-determined because the number of unknowns in $\widetilde{\boldsymbol{P}}_\mathrm{i}$, $\widetilde{\boldsymbol{O}}$, and $\widetilde{\boldsymbol{P}}_\mathrm{o}$ exceeds the available measurements in $\widetilde{\boldsymbol{R}}$. To obviate this, we exploit the fact that wave scattering in realistic media is concentrated near the main diagonal of the transmission matrix. We therefore constrain the optimization to the dominant off-diagonals corresponding to the strongest deflection angles (Fig. 3d). Choosing too few components neglects relevant scattering contributions, while too many reintroduce overfitting. To balance these effects, we introduce *Progressive Reconstruction of Off-diagonals* (PRO): the optimization begins with a small set of off-diagonals to establish a stable initialization and then progressively incorporates higher-angle components, as illustrated by Stages I–III in Fig. 3d.

We validated this reconstruction framework using numerically simulated reflection matrix (see Methods). The simulated sample comprised a volumetric scattering medium with an embedded object (Figs. 3a-c; Supplementary Fig.

4), a configuration in which conventional confocal imaging yielded a severely distorted image (Fig. 3e). By applying PRO and expanding the number of reconstructed diagonals from 9 to 121 (Fig. 3f), we recovered the correct structure with increasing fidelity, approaching the ideal image (Fig. 3g).

Fourier transforming the optimized transmission matrices $\widehat{\boldsymbol{P}}_{\mathrm{i}}$ and $\widehat{\boldsymbol{P}}_{\mathrm{o}}$ yields the spatially varying input and output PSFs, respectively. The correlations between the reconstructed and ground-truth PSFs exceed 80% on average in high-reflectivity regions (Supplementary Fig. 6). Direct optimization of all 121 off-diagonals without the progressive strategy led to poorer results (Supplementary Fig. 7) due to overfitting, confirming that PRO effectively regularizes the inverse problem by expanding from small to large scattering angles. Off-diagonals not yet included in the model act as effective multiple-scattering noise. As PRO advances, these components are successively incorporated, converting this noise into usable signal. Overall, the usable signal from multiply scattered light was about eight times higher than that from the ballistic component (Methods).

For comparison, we applied CLASS[4] and patch-CLASS[9] algorithms (see Supplementary Note 5) to the same simulated dataset. CLASS, which assumes a single isoplanatic patch over the full field of view, failed to recover the object because the estimated PSFs were poorly correlated with the ground truth (Supplementary Fig. 8a). Patch-CLASS, which assumes locally invariant PSFs within small patches, produced modest improvement but remained limited by discontinuities between adjacent patches and inaccurate PSFs (Supplementary Fig. 8b), in contrast to the smoothly varying PSFs recovered by our method. These limitations arise from the assumption of local shift invariance—an assumption that breaks down when the scattering medium is located near the object plane, giving rise to strong spatial variation in the PSFs. We also applied the Multiple Scattering Tracing (MST) algorithm[21], which models the medium as a stack of discrete phase-plate layers. MST failed to recover the structure as well (Supplementary Fig. 8c), highlighting its limitations in complex volumetric scattering media. The PSF correlation with the ground truth for these methods was less than 30%, far below the 80% achieved by PRO. Together, these results highlight the inherent limitations of patch-based and layer-based approximations and underscore the importance of PRO, which reconstructs continuously varying, position-dependent PSFs essential for accurate imaging in complex scattering environments.

### *In vivo* imaging of bone cells through an intact mouse skull

Osteocytes are the primary mechanosensory cells embedded deep within mineralized bone and constitute over 90% of all bone cells[25]. Each osteocyte extends dozens of long dendritic processes, slender projections that form a vast intercellular network critical for sensing mechanical load and coordinating bone remodeling[26,27]. Visualizing these fine structures *in vivo* has remained challenging due to strong multiple scattering within the dense bone matrix, which severely degrades image resolution. Conventional imaging techniques often fail to resolve even the cell bodies, let alone their fine dendritic processes. This limitation arises because the mouse skull acts as a volumetric scattering medium. The scattering bone matrix surrounds the target cells, and this proximity results in highly complex, spatially varying PSFs. Our method overcomes these limitations, enabling clear visualization of both osteocyte somas and their processes through intact skulls.

To demonstrate this capability, we applied our angular decomposition-based reconstruction framework *in vivo*. For a 20-week-old mouse with a skull thickness of ~200 μm, the scalp was removed and a circular glass coverslip was affixed to the exposed parietal bone using biocompatible adhesive (see Methods). Time-gated reflection matrices were acquired with the imaging objective focused inside the skull, targeting depths of 80 μm and 140 μm below the skull surface (blue dashed line in Fig. 4a). The field of view (FOV) was approximately 118 × 118 $\mu m^2$, and the illumination wavelength was 1.3 μm (see Methods for the full experimental setup).

Conventional confocal imaging yielded blurry, featureless images even at 80 μm depth (Fig. 4b). In contrast, PRO successfully recovered high-resolution images of osteocytes, revealing fine dendritic processes (Fig. 4c). To facilitate reconstruction, the FOV was divided into nine sub-regions (50 × 50 $\mu m^2$ each), which were reconstructed independently and then stitched to form the final image. This division was for computational convenience and unrelated to optical isoplanaticity, as PRO does not rely on local PSF invariance. Local reconstructions from the dashed boxes in Fig. 4c (Fig. 4d) confirm that individual osteocytes are clearly resolved after PRO. For comparison, we also applied patch-CLASS and MST to the same local regions. Patch-CLASS applies CLASS separately to multiple patches, while MST models the medium as a stack of discrete phase-plate layers. Their reconstructions (Fig. 4e,f) show that neither method is able to fully recover the fine cellular structure resolved by PRO.

At 140 μm depth, the same trend was observed: the confocal image failed to reveal discernible structures (Fig. 4g), whereas PRO restored detailed images of osteocytes across the FOV (Fig. 4h). Independent analysis of the boxed sub-regions once again verified accurate recovery of individual cells (Fig. 4i), while patch-CLASS and MST reconstructions of the same regions (Fig. 4j,k) failed to recover the fine structure (see a systematic comparison with existing approaches in Supplementary Fig. 9).

To quantify this improvement, we compared the predicted reflection matrices with the measured ones and found that PRO increased the usable signal by a factor of ten relative to that from the ballistic component (Methods). These findings collectively demonstrate the shortcomings of layer-based and patch-based models and emphasize the necessity of PRO for progressively converting multiply scattered light into usable signal, enabling visualization of structures otherwise invisible *in vivo*.

## Discussion

We have introduced the scattering-angle basis as a general physical framework for wave transport in complex scattering media. This representation links each scattering angle to a spatially invariant component of the point spread function, modulated only by a simple phase factor. The framework offers key advantages over existing layer-based models for describing light transport through volumetric, forward-scattering media such as biological tissues, where scattering occurs continuously at all depths and the medium's response is sparse in the angular domain.

To exploit this framework experimentally, we developed the Progressive Reconstruction of Off-diagonals (PRO) algorithm, which iteratively incorporates larger scattering-angle components. By progressively expanding the accessible angular bandwidth, PRO converts multiply scattered light—traditionally treated as stochastic noise—into coherent information that sharpens the reconstructed images. PRO successfully recovered fine osteocyte structures through intact mouse skull. This regime represents a volumetric scattering medium where the distribution of scatterers near the object plane makes layer-based and patch-based models ineffective.

Despite these capabilities, the scattering-angle approach has room for improvement and expansion. Although PRO robustly reconstructs up to 121 scattering-angle components, retrieving contributions from larger deflection angles remains challenging. As the angular spread induced by the medium broadens, the relative weight of high-angle components increases, and if these become substantial, reconstruction fidelity deteriorates (Supplementary Fig. 10). Future algorithmic developments will be needed to handle broader angular spreads more effectively. In principle, our framework can extend to nonlinear wave-scattering modalities such as second-harmonic generation[28,29] and coherent anti-Stokes Raman scattering. Moreover, in layer-based models, physical aberration correction can be implemented by conjugating a spatial light modulator to an identified scattering layer. In contrast, scattering-angle components do not correspond to a single spatial plane, so realizing the physical inverse of our model for real-time wavefront shaping will require the design and development of new hardware architectures capable of directly modulating angular modes.

From the perspective of solving inverse scattering problems, the scattering-angle basis offers a new paradigm in parameter efficiency. A critical factor in inverse scattering is the optimal parametrization of the wave-propagation model. In scenarios such as through-skull deep-brain imaging, where the primary scattering layer (the skull) is far from the target neurons, the scattering potential is sparse along the depth axis; in such "depth-wise sparse" regimes, layer-based models are effective[21,22]. However, when imaging targets embedded within the skull itself, accurately modeling the continuously varying bone matrix requires a densely packed series of phase plates, leading to severe over-parameterization and unstable convergence. In contrast, the scattering-angle basis exploits the angular sparsity of forward-scattering media, allowing complex volumetric scattering to be described with a relatively small number of angular parameters.

Collectively, these observations suggest that the scattering-angle framework and conventional layer-based models are complementary. The optimal choice depends on the specific scattering characteristics of the medium—whether it is primarily depth-wise sparse or angularly sparse. In the long term, a synergistic combination of both approaches could capture a broader range of multiple-scattering events. Because the achievable imaging depth is fundamentally dictated by the extent to which multiple scattering can be inverted, the scattering-angle framework represents a significant step toward pushing the physical limits of deep optical imaging.

## Methods

### *Simulation of transmission and reflection matrices*

Let the size of ROI be $L \times L$. If the wavelength of light is $\lambda$, and numerical aperture of objective is $\alpha$, then the cut-off frequency is $k_\mathrm{c} = 2\pi\alpha/\lambda$, so that $|\mathbf{k}_\mathrm{i}| \leq k_\mathrm{c}$. The spatial frequency sampling is $\delta k = 2\pi/L$. In the spatial domain, the resolution is $\delta r = \lambda/2\alpha$, consistent with the Nyquist sampling criterion for the cut-off bandwidth. The transmission matrix $\widetilde{\boldsymbol{P}}_{\mathrm{i}}$, represented in the scattering-angle basis, comprises multiple off-diagonals, characterized by different sub-diagonal matrix $\widetilde{\boldsymbol{Q}}_{\mathrm{i}}^{(\Delta\mathbf{k})}$. Here, $\Delta\mathbf{k} = (n_x\delta k, n_y\delta k)$, where $n_x$ and $n_y$ are integers. The position-dependent PSFs are obtained from the Fourier transform of $\widetilde{\boldsymbol{P}}_{\mathrm{i}}$; similarly, the PSFs corresponding to each scattering angle ($\Delta\mathbf{k}$) can be obtained from $\widetilde{\boldsymbol{Q}}_{\mathrm{i}}^{(\Delta\mathbf{k})}$. The matrix elements were assigned either randomly or following a specific distribution. For the simulations in Fig. 2, we used two types of matrices: one for the special case of a single off-diagonal, and another in which the off-diagonals followed a Gaussian amplitude distribution centered on the main diagonal. The parameters were $\lambda = 1$ µm and $\alpha = 1$ for an ROI of $15 \times 15$ µm$^2$, giving $\delta r = 0.5$ µm and $\delta k = 0.4054$ µm$^{-1}$.

To simulate the reflection matrix in Fig. 3, we first constructed the transmission matrix of a volumetric scattering medium. The medium was modelled as a stack of phase plates separated by $\lambda/2\alpha$ in the $z$ direction, with a transverse ($x$-$y$) grid spacing of $\lambda/2\alpha$. The first and last phase plates were assigned random phase maps, while the intermediate phase plates were generated as weighted interpolations of the two boundary layers, producing a smooth refractive-index variation. Additionally, small random refractive-index perturbations were added to each layer to mimic realistic stochastic variations across the 3D volume. The input transmission matrix $\widetilde{\boldsymbol{P}}_{\mathrm{i}}$ was computed using angular spectrum propagation, where waves propagated freely between the layers and each layer contributed a phase retardation determined by the refractive index variation[30]. The reflection matrix $\widetilde{\boldsymbol{R}}$ was obtained as $\widetilde{\boldsymbol{R}} = \widetilde{\boldsymbol{P}}_{\mathrm{o}}\widetilde{\boldsymbol{O}}\widetilde{\boldsymbol{P}}_{\mathrm{i}}$, where $\widetilde{\boldsymbol{P}}_{\mathrm{o}}$ is the transpose of $\widetilde{\boldsymbol{P}}_{\mathrm{i}}$ and represents the wavefront distortion in the output path. For these simulations, we used $\lambda = 1.3$ µm and $\alpha = 1$ over an ROI of $46 \times 46$ µm$^2$, which corresponds to $\delta r = 0.65$ µm and $\delta k = 0.1361$ µm$^{-1}$. Notably, the main diagonal ($\Delta\mathbf{k} = 0$), which contributes to the spatially invariant part of the PSF, accounted for only 14% of the total energy, while 121 diagonals collectively accounted for over 75% of the transmitted energy for each pathway, resulting in strongly position-dependent PSFs (Supplementary Fig. 5).

***Pearson correlation***

The loss function uses the Pearson correlation coefficient between the columns of the predicted reflection matrix (based on the parameters being optimized) and the ground truth. Each column represents the complex-field image at the detector corresponding to a particular input channel. For two column vectors $X_a$ and $X_b$, the Pearson correlation coefficient is defined as

$$\mathrm{PC}(X_a,\ X_b) = \frac{\overline{(X_a - \overline{X_a})(X_b - \overline{X_b})^*}}{\sqrt{\overline{|X_a - \overline{X_a}|^2}}\sqrt{\overline{|X_b - \overline{X_b}|^2}}} \tag{6}$$

where $\overline{(\cdot)}$ denotes the mean of the enclosed expression, and $(\cdot)^*$ represents the complex conjugate. Here, the numerator is the covariance between $X_a$ and $X_b$, and the denominator is the product of their standard deviations.

The Pearson correlation is also used to quantify how much deterministic signal is recovered from multiply scattered light. To do this, we computed the correlation between the PRO-predicted reflection matrix and the measured reflection matrix for two cases: (i) reconstruction using only the main diagonal (ballistic component) and (ii) reconstruction using all off-diagonals incorporated by PRO. For each case, the correlation was computed column-wise and averaged across all channels. The ratio of these two correlations defines the relative increase in usable signal provided by PRO. In the simulations, the main diagonal carried ~14% of the transmitted energy, whereas the full set of 121 diagonals accounted for ~75%, corresponding to an eightfold increase in matrix correlation. In the *in vivo* experiments, this correlation ratio was ~10, indicating that PRO enabled ten times more deterministic signal to be recovered compared to the ballistic case.

Note that when analyzing our results, including the correlation between PSFs or between reflection matrices, we report the magnitude of the computed correlation coefficient.

***Optimization procedure***

By expanding the transmission matrices in the scattering-angle basis, we minimize the loss function defined in Eq. (5):

$$\mathcal{L} = -\mathrm{PC}\big(\widetilde{\boldsymbol{R}},\ \widehat{\boldsymbol{P}}_{\mathrm{o}}\widehat{\boldsymbol{O}}\,\widehat{\boldsymbol{P}}_{\mathrm{i}}\big) = -\mathrm{PC}\left[\widetilde{\boldsymbol{R}},\ \sum_{\Delta\mathbf{k}'}\widehat{\boldsymbol{Q}}_{\mathrm{o}}^{(\Delta\mathbf{k}')}\widehat{\boldsymbol{O}}\sum_{\Delta\mathbf{k}}\widehat{\boldsymbol{Q}}_{\mathrm{i}}^{(\Delta\mathbf{k})}\right] \tag{7}$$

Here, $\widehat{\boldsymbol{Q}}_{\mathrm{i}}^{(\Delta\mathbf{k})}$ and $\widehat{\boldsymbol{Q}}_{\mathrm{o}}^{(\Delta\mathbf{k}')}$ are the various off-diagonals comprising $\widehat{\boldsymbol{P}}_{\mathrm{i}}$ and $\widehat{\boldsymbol{P}}_{\mathrm{o}}$. The ranges of summation over $\Delta\mathbf{k} = \mathbf{k} - \mathbf{k}_{\mathrm{i}}$ and $\Delta\mathbf{k}' = \mathbf{k}_{\mathrm{o}} - \mathbf{k}'$ were progressively expanded across optimization stages, as shown in Fig. 3d. The optimization was carried out using the ADAM optimizer with an empirically determined initial learning rate of 0.01. Training was conducted for 500 epochs with a cosine annealing schedule that gradually decayed the learning rate to 1% of its initial value. The optimization was performed in batches: for each batch, predicted images were generated for a random subset of input illuminations, and the Pearson correlation loss was computed against the ground truth on a column-by-column basis. The model parameters were subsequently updated based on this loss. Since the calculated Pearson correlation is complex, we used the mean of its real and imaginary parts as the loss function for the optimization.

The object was optimized in the spatial domain at a sampling resolution of $\lambda/4\alpha$, and then Fourier transformed into the spatial-frequency domain $\widehat{\boldsymbol{O}}$ for use in the loss function. To suppress high-frequency artifacts, a circular Tukey window was applied in the Fourier domain, defined as

$$w(\rho) = \begin{cases} 1, \quad 0 \le \rho \le \gamma \\ 1/2(1 + \cos(\pi(\rho - \gamma)/\gamma), \quad \gamma < \rho \le 1 \\ 0, \quad \rho > 1 \end{cases} \tag{8}$$

Here, $\rho$ is the normalized radial distance from the image center, and $\gamma$ controls the width of the transition region. After training, the optimized object was filtered using a Tukey window with $\gamma = 0.5$.

In each batch, all elements of $\widehat{\boldsymbol{P}}_{\mathrm{o}} = \sum_{\Delta\mathbf{k}'} \widehat{\boldsymbol{Q}}_{\mathrm{o}}^{(\Delta\mathbf{k}')}$ are updated because they affect every prediction. In contrast, for $\widehat{\boldsymbol{P}}_{\mathrm{i}} = \sum_{\Delta\mathbf{k}} \widehat{\boldsymbol{Q}}_{\mathrm{i}}^{(\Delta\mathbf{k})}$, only the columns corresponding to the input illuminations for that batch receive gradient updates. The additional frequency of updates leads to more accurate convergence of the output transfer functions. Furthermore, owing to its position at the front of the matrix chain, $\widehat{\boldsymbol{P}}_{\mathrm{o}}$ receives more stable and well-conditioned gradients during optimization, while $\widehat{\boldsymbol{P}}_{\mathrm{i}}$ is more susceptible to gradient distortion due to appearing downstream, further limiting its reconstruction accuracy.

The algorithm was implemented in Python (v3.11.13, Anaconda distribution) using PyTorch (v2.9, CUDA 12.8) on a personal workstation (OS: Microsoft Windows 11 Pro; CPU: Intel Core Ultra 7 265K; RAM: 192 GB; GPU: NVIDIA RTX 6000 Pro Blackwell, 96 GB VRAM). Data analysis was performed in MATLAB (v2025a). Exemplary code for implementation of the PRO workflow is publicly available[31].

***Progressive Reconstruction of Off-diagonals (PRO)***

Here we detail the specific off-diagonals targeted during the implementation of PRO. The transverse components of the scattering angles are $\Delta\mathbf{k} = (n_x\delta k, n_y\delta k)$ and $\Delta\mathbf{k}' = (n'_x\delta k, n'_y\delta k)$. In the first stage, we target diagonals with $n_x, n_y, n'_x, n'_y \in [-1, 1]$, corresponding to 9 adjacent diagonals in both $\widehat{\boldsymbol{P}}_{\mathrm{o}}$ and $\widehat{\boldsymbol{P}}_{\mathrm{i}}$. These off-diagonals are symmetrically distributed around the main diagonal ($\Delta\mathbf{k}, \Delta\mathbf{k}' = 0$). The first stage is initialized using the results of patch-CLASS (see Supplementary Note 6 and Supplementary Fig. 11 for details). The results of the first stage are used to initialize the second stage, where $n_x, n_y, n'_x, n'_y \in [-2, 2]$, targeting 25 diagonals. In general, for the $N^{\mathrm{th}}$ stage, the optimization targets all off-diagonals corresponding to $n_x, n_y, n'_x, n'_y \in [-N, N]$, such that the number of targeted diagonals is $(2N + 1)^2$.

As elucidated in the main text, this progressive approach gradually refines both the PSFs and the object, yielding more accurate results than optimizing a large number of diagonals simultaneously. To illustrate this, the number of channels in the input/output transmission matrix is $\left(\frac{2k_c}{\delta k}\right)^2 = \left(\frac{L}{\delta r}\right)^2$, but due to the limited numerical aperture, the effective number of channels in each diagonal is approximately $N_c \approx \frac{\pi}{4}\left(\frac{2k_c}{\delta k}\right)^2$. For the simulation results discussed in Fig. 3, the ROI spans $71\delta r \times 71\delta r$. If the optimization targeted only the main diagonal—treating the entire ROI as a single isoplanatic patch—the number of parameters would be just 3853 for each pathway. In contrast, accounting for 121 off-diagonals increases this to 455961 parameters, underscoring the method's complexity. The experimental results shown in Fig. 4 use a similar ROI size and therefore involve the same number of unknown parameters. Each stage typically runs for 500 epochs; for ROI spanning $71\delta r \times 71\delta r$, each epoch requires ~100 ms, leading to per-stage runtime of ~50 s.

As discussed above, the target diagonals selected at each stage of PRO are chosen based on symmetry. However, the target diagonals can be adjusted without loss of generalization. In principle, PRO could extend to arbitrarily large scattering angles. However, practical limitations arise from noise in experimentally measured data, including reflections from non-target layers and multiple-scattering backgrounds. Further, while increasing the number of targeted off-diagonals reduces the relative fraction of untargeted off-diagonals that act as noise, it simultaneously increases the number of optimization parameters, raising the risk of overfitting. In addition, the signal from newly targeted off-diagonals may be too weak to be recovered reliably. Consequently, improvement in image quality at each stage is accompanied by an increase in random noise and artifacts as well. This creates a trade-off between recovering additional off-diagonals and maintaining acceptable noise levels when applying PRO to experimental data. In this study, PRO was terminated when additional stages no longer improved image quality due to noise and overfitting; empirically, this occurred at the fifth stage.

***Experimental acquisition of reflection matrix***

For the measurement of the reflection matrix, we utilized laser scanning reflection-matrix microscopy system based on an interferometric confocal reflectance microscope (details of the setup can be found in our previous work[21]). For imaging a 20-week-old mouse with intact skull, we used a pulsed laser (INSIGHT X3, Spectra Physics, 1.3 µm wavelength, 19 nm bandwidth) which provided a coherence-gated window of ~25 µm optical path length. The laser beam was split into sample and reference beams at a beam splitter and recombined by another beam splitter to form a Mach–Zehnder interferometer. The sample beam was relayed through two galvanometer mirrors and focused onto the skull using an objective lens (Olympus, 25×, NA 1.05). The backscattered signal was descanned by the galvanometer mirrors and recorded by a high-speed camera (InGaAs, Cheetah800, Xenics, 6.8 kHz frame rate) using off-axis digital holography. The focused spot was scanned at steps of $\delta r = \lambda/2\alpha$, consistent with the cut-off frequency of $k_c = 2\pi\alpha/\lambda$ ($\alpha$ is the numerical aperture). The acquired complex-field images were used to construct the time-gated reflection matrix $\boldsymbol{R}$ in the spatial domain, with the input channels ($\mathbf{r}_i$) along columns and the output channels ($\mathbf{r}_o$) along rows. Subsequently, $\boldsymbol{R}$ was Fourier transformed to obtain the angle-to-angle reflection matrix $\widetilde{\boldsymbol{R}}$. For the FOV of 118 × 118 µm² in our experiments, the data acquisition time was about 5 s.

***Preparation of samples for in vivo imaging***

All animal procedures followed ethical guidelines and were approved by the Korea University Institutional Animal Care and Use Committee (KUIACUC-2022-0013). The mouse was housed in temperature-controlled (20–22 °C) and humidity-controlled (50–55%) facilities under a 12-h light/12-h dark cycle. The intact skull window preparation followed previously reported procedure[21]. Mouse was anesthetized with isoflurane (1.5–2% in oxygen, breathing rate ~1 Hz), and the body temperature was maintained at 37–38 °C using a heating blanket. During surgery and imaging, the eyes were protected with ophthalmic ointment. After hair removal using Nair, the scalp was excised to expose the parietal bones, and any residual connective tissue was carefully removed using sterile forceps. A round glass coverslip (#1, Warner Instruments; diameter: 5 mm; thickness: ~100 µm) was affixed to the skull using ultraviolet-curable adhesive (Loctite 4305). A custom metal plate was then attached to the skull with cyanoacrylate glue for head fixation, and the exposed area was sealed with dental cement (Dentsply DeTrey GmbH, Germany). Postoperative care included intramuscular injection of dexamethasone (1 mg/kg) to reduce inflammation at the surgical site. During imaging sessions, the mouse was maintained under a steady flow of isoflurane (1.2–1.5%, breathing rate ~1.5–2 Hz), and body temperature was kept at 37–38 °C using a heating blanket on a 3D motorized stage.

**Funding**

National Research Foundation of Korea (NRF) grant funded by the Korea (MSIT) (RS-2024-00442818, RS-2025-24132968, RS-2025-24523003, NRF-2021R1C1C2008158 and RS-2023-00251628); POSCO Science Fellowship of POSCO TJ Park Foundation.

**Author contributions**

W.C., Y.-R.L., S.K., and K.G. conceived the project. K.G. implemented the framework and algorithm, and analyzed the experimental data, with assistance from D.H.T. and Y.K., S.K., S.Y. conducted the experiments, and J.-H.H. prepared the biological samples. K.G., Y.-R.L., and W.C. wrote the manuscript, and all authors contributed to finalizing it. W.C. supervised the project.

**Data Availability**

The simulated data presented in Fig. 3 were generated and analyzed using publicly available code[31]. The experimental datasets supporting Fig. 4 are not publicly available due to large data size, but will be provided by the corresponding author upon reasonable request.

**Code Availability**

The code for generation and analysis of the simulated data in Fig. 3 are publicly available in the following figshare repository: https://doi.org/10.6084/M9.FIGSHARE.28953386[31]. The code for analysis of the experimental data is not publicly available, but will be made available on request.

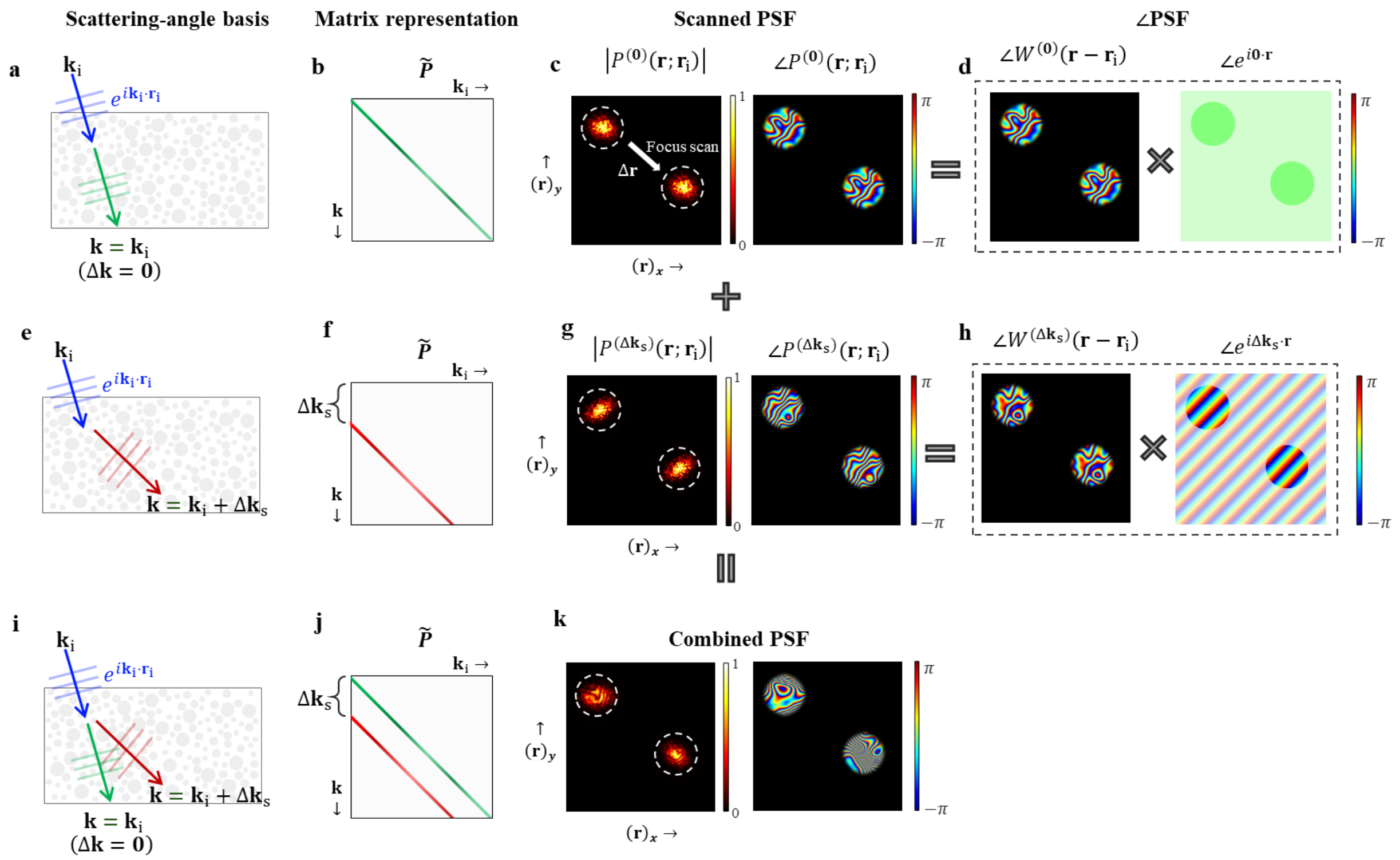

**Fig. 1. Scattering angle representation of position-variant PSF. a**, An incident plane wave with transverse wavevector $\mathbf{k}_\mathrm{i}$ interacts with the scattering medium, producing transmitted waves with wavevectors $\mathbf{k}$. Here, the undeviated (ballistic) component corresponding to $\Delta\mathbf{k} = \mathbf{0}$ is shown. **b**, In this case, the transmission matrix $\tilde{\boldsymbol{P}}$ contains only the diagonal component (green line). **c**, The ballistic PSFs corresponding to two illumination positions, $\mathbf{r}_\mathrm{i}$ and $\mathbf{r}_\mathrm{i} + \Delta\mathbf{r}$, are shown in their amplitude (left panel) and phase (right panel). **d**, The phase of the ballistic PSF, which is spatially invariant and equal to that of $W^{(\mathbf{0})}(\mathbf{r} - \mathbf{r}_\mathrm{i})$. **e**, The case when each incident wave results in a scattered wave corresponding to $\mathbf{k} = \mathbf{k}_\mathrm{i} + \Delta\mathbf{k}_\mathrm{s}$. **f**, The transmission matrix now contains a single off-diagonal component (red line). **g**, The resulting PSF is spatially invariant in amplitude, but has an overall phase shift. **h**, Phase of the PSF in (**g**), consisting of spatially invariant $W^{(\Delta\mathbf{k}_\mathrm{s})}(\mathbf{r} - \mathbf{r}_\mathrm{i})$ (left panel) modulated by the spatially varying oscillatory phase term, $e^{i\Delta\mathbf{k}_\mathrm{s}\cdot\mathbf{r}}$ (right panel). **i**, The case when the scattering medium gives rise to two distinct scattering angles, $\Delta\mathbf{k} = 0$ (green) and $\Delta\mathbf{k} = \Delta\mathbf{k}_\mathrm{s}$ (red). **j**, Consequently, $\tilde{\boldsymbol{P}}$ comprises a diagonal component (green line) and a single off-diagonal component (red line). **k**, The two PSFs due to the ballistic and the scattered components superimpose to create a spatially varying PSF in both amplitude and phase. Color bars for amplitude maps: normalized amplitude. Color bars for phase maps: phase in radians.

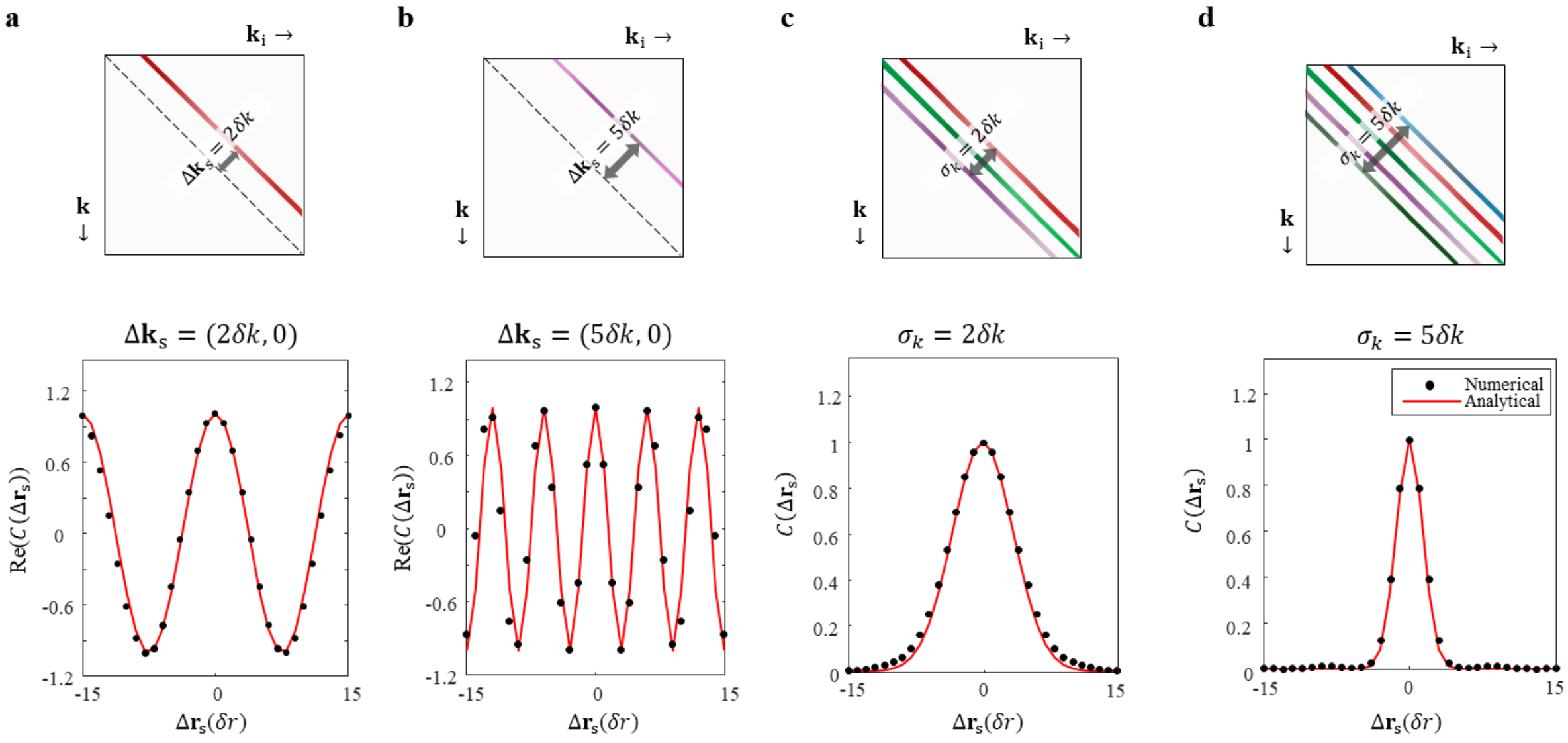

**Fig. 2. Translational correlation of PSFs for different types of scattering media. a**, The scattering medium generates a single wave deflection at $\Delta\mathbf{k}_s = (2\delta k, 0)$. The corresponding transmission matrix (top) contains a single off-diagonal. The translational correlation $C(\Delta\mathbf{r}_s)$ (bottom) is oscillatory, with the oscillation frequency given by $\Delta\mathbf{k}_s$. Since $C(\Delta\mathbf{r}_s)$ is complex, the real part is plotted for better visualization. Black markers: numerically obtained from transmission matrix. Red curve: analytical prediction. **b**, Same as (**a**), but for the case of $\Delta\mathbf{k}_s = (5\delta k, 0)$. **c**, The scattering medium generates a Gaussian spatial frequency spectral broadening with $\sigma_k = 2\delta k$ (top). The corresponding translational correlation (bottom) is a Gaussian with standard deviation equal to $\sqrt{2}/\sigma_k$. **d**, Same as (**c**), but when $\sigma_k = 5\delta k$. $\Delta\mathbf{r}_s$ is in the units of $\delta r$; $\delta k$ and $\delta r$ are the angular and spatial resolutions in the spatial frequency domain and the spatial domain, respectively.

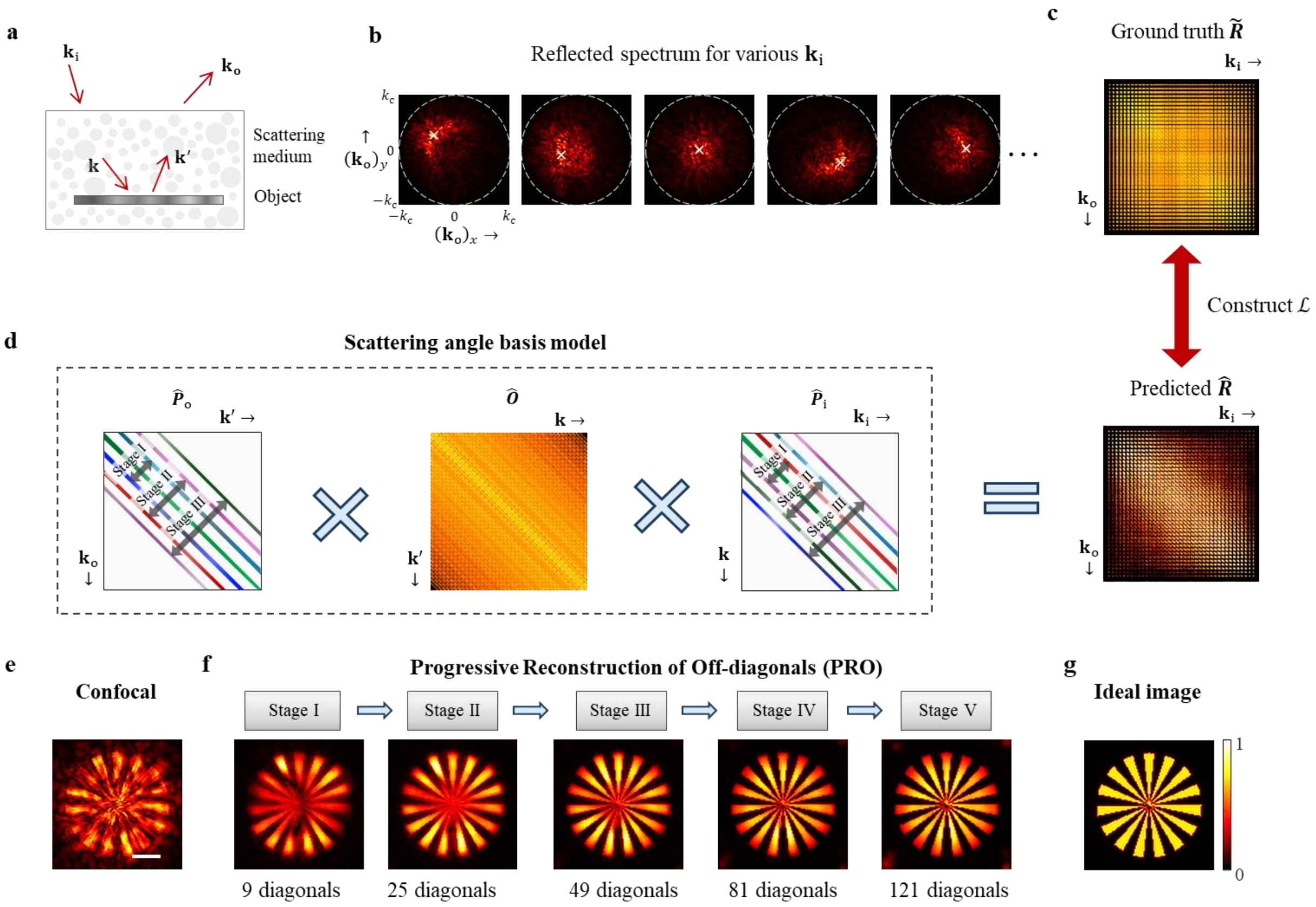

**Fig. 3. Implementation of image reconstruction in the scattering-angle basis.** **a**, Schematic of reflection-mode imaging configuration in the scattering-angle basis. Scattering trajectories of both incident and reflected light are illustrated to emphasize the complexity of angular deflections within a volumetric medium. **b**, Representative reflected spectra $\tilde{E}(\mathbf{k}_o; \mathbf{k}_i)$ measured for different incident wavevectors $\mathbf{k}_i$. The center of each spectrum corresponds to $\mathbf{k}_o = (0,0)$; the × mark indicates $\mathbf{k}_o = \mathbf{k}_i$. The dashed circles indicate bandwidths with radius $k_c = 2\pi\alpha/\lambda$, where $k_c$ is the cutoff spatial frequency, $\lambda$ is the wavelength, and $\alpha$ is the numerical aperture. **c**, Simulated ground-truth reflection matrix $\tilde{\boldsymbol{R}}$ constructed from the reflected spectra in (**b**). **d**, Schematic construction of the predicted reflection matrix in the scattering-angle basis. The input and output transmission matrices contain only a limited number of dominant off-diagonals corresponding to specific angular deflection components. **e**, Conventional confocal image reconstructed from the simulated reflection matrix, showing severe distortion caused by multiple scattering. **f**, Object reconstructions obtained using the proposed progressive reconstruction of off-diagonals (PRO) algorithm, as the number of optimized off-diagonals increases from 9 to 121. **g**, Ideal confocal image obtained in the absence of scattering medium. Scale bar in (**e**), 10 μm (applies to (**f**) and (**g**)); all images are displayed on the same scale.

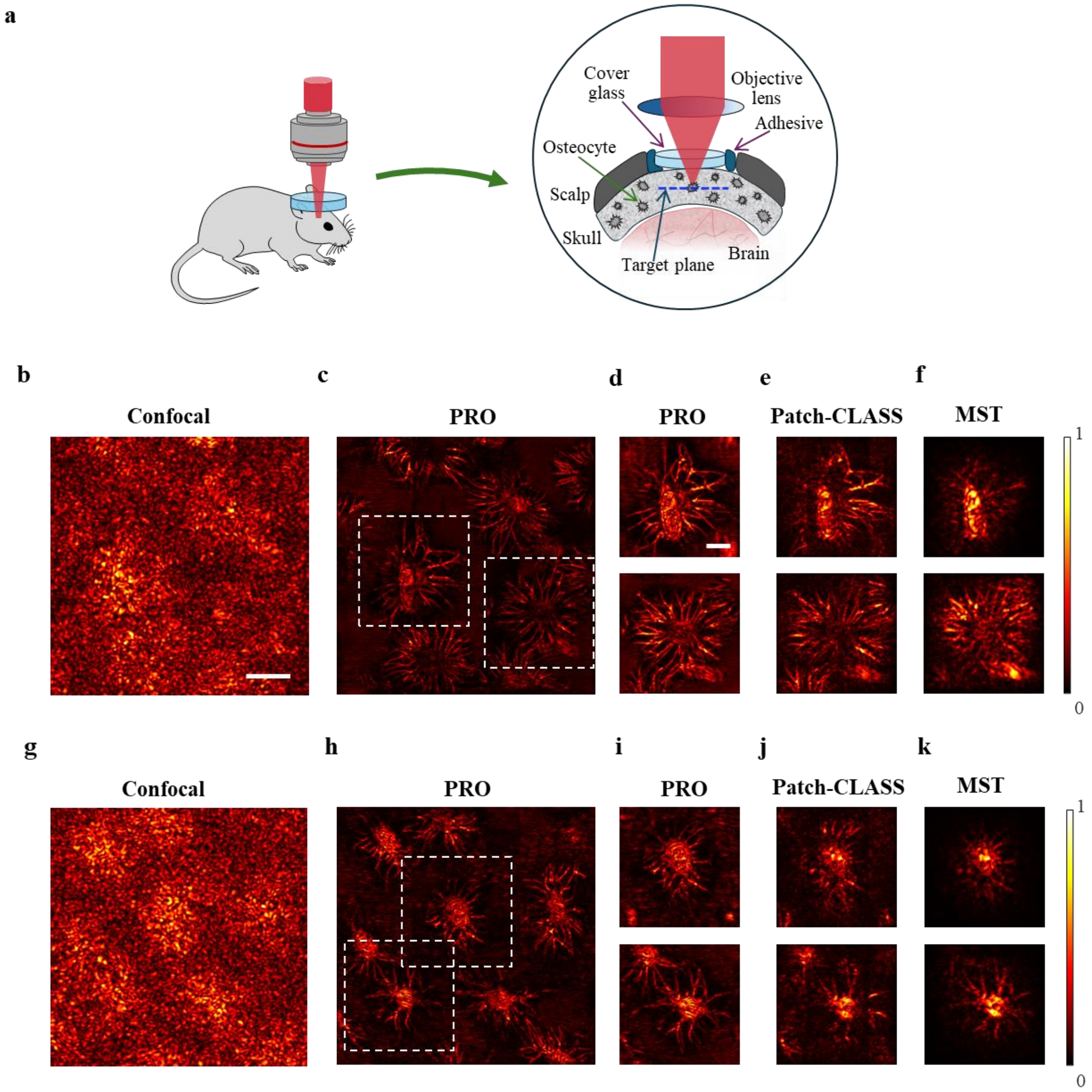

**Fig. 4. *In vivo* imaging inside mouse skull bone using the scattering-angle framework. a**, Schematic of the experimental configuration. Both the focused illumination and the time-gating window are set to a target plane within the skull. **b**, Confocal reflectance image acquired at a depth of 80 μm from the skull surface of a 20-week-old living mouse (skull thickness ~200 μm). The image is severely degraded due to volumetric scattering. **c**, High-resolution image of osteocytes recovered using the proposed PRO method, revealing fine structural details otherwise obscured in (**b**). The field of view was divided into 9 regions (50 × 50 μm² each), which were reconstructed independently and subsequently stitched to form the final image. **d**, Independently reconstructed osteocyte images from the local regions marked by dashed boxes in (**c**). **e**, Patch-CLASS reconstructions of the same local regions as in (**d**). **f**, MST reconstructions of the same regions as in (**d**). **g-k**, Same as (**b**)-(**f**), but acquired at a depth of 140 μm beneath the skull surface. At this depth, the confocal image (**g**) fails to resolve the osteocytes, whereas PRO successfully recovers fine structural details (**h**, **i**), outperforming patch-CLASS and MST (**j**, **k**). All images are normalized with respect to their peak values. Scale bar in (**b**), 20 μm (applies to (**c**, **g**, **h**)); scale bar in (**d**), 10 μm (applies to (**e**, **f**, **i-k**)).